\documentclass[a4paper]{jpconf}
\usepackage{graphicx}
\usepackage{epstopdf}
\usepackage{amsmath,amssymb}
\usepackage{multirow}
\usepackage{color}
\usepackage[normalem]{ulem}
\renewcommand{\d}{{\rm d}}
\newcommand{\imai}{{\rm i}}

\graphicspath{ {images/} }
 
\begin{document}
\title{Simple electron-electron scattering in non-equilibrium Green's function simulations}

\author{David O. Winge$^1$, Martin Francki\'{e}$^1$, Claudio Verdozzi$^1$, Andreas Wacker$^1$ and Mauro F. Pereira$^2$}

\address{$^1$ Division of Mathematical Physics, Department of Physics, Lund University, Box~118, SE-221~00~Lund, Sweden}
\address{$^2$ Materials and Engineering Research Institute, Sheffield Hallam University, City Campus, Howard Street, Sheffield S1~1WB, UK}

\ead{david.winge@teorfys.lu.se}

\begin{abstract}
In this work we include electron-electron interaction beyond
Hartree-Fock level in our non-equilibrium Green's function approach by
a crude form of GW through the
Single Plasmon Pole Approximation. This is achieved by treating all
conduction band electrons as a single effective band screening the
Coulomb potential. We describe the corresponding self-energies in this
scheme for a multi-subband system. In order to apply the formalism to
heterostructures we discuss the screening and plasmon dispersion in
both 2D and 3D systems. Results are shown for a four well quantum
cascade laser with different doping concentration where comparisons to
experimental findings can be made. 
\vspace{3mm} \\
\textit{Accepted for publication in \textbf{Journal of Physics: Conference Series} by IOP Publishing Ltd.,
as a proceeding of the PNGF6 workshop in Lund, Sweden, 2015.}
\end{abstract}

\section{Introduction}
Since the quantum cascade laser (QCL) was introduced more than 20 years ago
\cite{FaistScience1994} it
has continuously been improved and redesigned to operate from the mid-infrared
all the way down to the Terahertz (THz) range. Compact laser
sources at these wavelengths are valuable for spectroscopic applications~ 
\cite{WilliamsNatPhotonics2007}
but a major problem is that THz QCLs do not operate at room temperature. 
This can be overcome via difference frequency generation 
which recently has been demonstrated at powers in the milli-watt range~\cite{BelkinPhysScripta2015}. 
Direct THz-QCLs still generate a lot of interest
in the research community due to the promise of higher wall-plug efficiency and the 
prospects of miniaturization of cryo-coolers.

The main temperature degrading mechanism of the THz QCLs is currently not fully 
understood and is still debated by the community~\cite{KhanalJOpt2014}. 
This is a challenge to theory and there is a great need of realistic
modeling tools that are able to treat all important quantum effects on the
same footing. A summary of different methods for simulating these structures
is found in~\cite{JirauschekApplPhysRev2014}.
Monte Carlo simulations have shown that electron-electron interaction, 
beyond the meanfield or Hartree approximation can
influence the dynamics of THz QCLs~\cite{CallebautAPL2004,ManentiJComputElectron2003,JirauschekPSSC2008}.

In this work we include and study the effects of a simple 
electron-electron scattering via the Single Plasmon Pole Approximation (SPPA)~
\cite{LundqvistPhysKondensMater1967,AndoRMP1982}. In this approximation we
capture both the static limit as well as dynamic effects. This gives an energy
dependent (non-local in time) interaction beyond the Hartree-Fock approximation.  
This has been studied in a similar model with promising results~\cite{SchmielauAPL2009}, 
and with this work we want to adapt the idea into our model described in 
Ref.~\cite{WackerQuantEl2013}. 
In other methods based on Non-Equilibrium Green's Functions (NEGF) and applied to
QCLs, electron-electron scattering has 
previously been implemented in a low order $GW_0$ approximation employing a purely
static screening model~\cite{KubisPRB2009}.

To simulate the periodic structure of the QCL, 
we use a product space of plane waves and confined states.
The Green's function evaluation provides us with the energy resolution of the states,
labeled by state and $k$ index, at the cost of inverting the self-energy matrices
at each $k$ and energy grid point. A number of elastic scattering processes as
well as longitudinal optical phonon scattering is treated in this formalism at 
a high computational cost. Adding a many-body term fully dependent on the polarization
would increase the complexity by one order, a formidable task in this
model, leading us to consider simpler schemes as done in this work.

This paper is organized as follows: first the theory of the SPPA is 
discussed and we show how it is implemented in the context of our
formalism. Then the dispersion of the plasmons is discussed as well as 
the exchange shift. In the last part a test structure is used to 
evaluate the effect of the inclusion of the new scattering mechanism.
Finally, we present the conclusions of our study.

\section{Theory}
In THz quantum cascade simulations, the relevant sources of scattering, 
beside electrons, are interface roughness, impurities from donor atoms,
acoustic and optical phonons as well as alloy composition.
All these effects have already been implemented and details can be 
found in Ref.~\cite{WackerQuantEl2013}.

The model discussed here employs a non-equilibrium Green's function
formalism where the system observables are extracted from the retarded
and lesser parts of the Green's function in the formalism of 
\cite{KeldyshSovPhysJETP1965,KadanoffBaymBook1962}.
These are solved using 
self-energies for the part of the Hamiltonian that is non-diagonal
in $\mathbf{k}$ and solved through the Dyson equation and the Keldysh
relation.
  
In this section we will describe how we formulate the self-energy
expressions for the screened interaction in the single plasmon pole
approximation (SPPA). First we will discuss how the conduction band is
treated as an effective band in order to apply the named
approximation. Then we will describe how we formulate and also
evaluate the Coulomb matrix elements. After this we acquire the lesser
and retarded part of the screened interaction before moving on to the actual
construction of the self-energies. This is done in close
correspondence to the treatment of phonons in
Ref.~\cite{WackerPhysRep2002} in order to keep the length of this
section reasonable. To this treatment we will here add the necessary
exchange part. Before we move on to the next section where we test the
model, we will also describe how the plasmon dispersion is treated.

\subsection{Screening from one effective band}
In frequency space the retarded
screened potential is given by its own Dyson equation, given here
as a matrix equation:
\begin{align}
\mathbf{W}(\mathbf{q},E)&=\mathbf{V(q)} + \mathbf{V(q)} 
\mathbf{P(q},E) \mathbf{W} (\mathbf{q},E) \\
\Longleftrightarrow \mathbf{W} (\mathbf{q},E) &= 
\underbrace{\left[1-\mathbf{V(q)}\mathbf{P}(\mathbf{q},E)\right]^{-1}}_{\epsilon(\mathbf{q},E)^{-1}} \mathbf{V(q)}.
\end{align}
where the $\mathbf{V}(\mathbf{q})$ is the bare interaction as a function of 
the momentum wave vector, described in more detail
below, and  $\mathbf{P}(\mathbf{q},E)$ is the polarization as a function
of momentum and energy $E$.
If we now assume an isotropic screening medium and that the
dielectric function $\epsilon(\mathbf{q},E)$ is not level
dependent, we can write the approximated retarded screened potential as
\begin{align} \label{Wret}
W_{ijkl}^{\rm ret}(\mathbf{q},E) = \frac{V_{ijkl}(q)}{\epsilon(\mathbf{q},E)}
\end{align}
where $V_{ijkl}(q)$ now holds the matrix structure. We consider the dielectric
function $\epsilon(\mathbf{q},E)$ given by the Lindhard formula~\cite{HaugKochBook2004}
which is retarded due to poles in the lower half of the 
complex plane. While we assume an effective band for the screening, all
other physical mechanisms are treated with full state dependence.

In this work we will assume a plasmon distribution in equilibrium with 
respect to the electron temperature. This approximation will ensure that
energy is not dissipated from the distribution of non-equilibrium electrons. 
In addition, this assumption will allow us to link the lesser part of the screened interaction
to the retarded through the fluctuation dissipation theorem~\cite{KadanoffBaymBook1962} as
\begin{align} \label{Wless_fd}
\mathbf{W^{<}}(\mathbf{q},E) &= -\imai f(E) \mathbf{A}(\mathbf{q},E) = - \imai f(E) \left[ -2\Im \{\mathbf{W}^{\rm ret} \} \right] % change 12 feb
 % (\mathbf{W^{\rm ret}}(\mathbf{q},E)-\mathbf{W^{adv}}(\mathbf{q},E)) \right]
\end{align}
where $f(E)$ is the (bosonic) plasmonic distribution function and $\mathbf{A}(\mathbf{q},E)$ is the spectral function, which can be written in terms of the retarded
Green's function leading to the rightmost expression. The electron temperature will be represented as the expectation
value $\langle E_k \rangle$ for the effective conduction band. This is with respect to the discrete energy for each 
state $\alpha$ as the single particle energies are $E_\alpha+E_k$. We calculate $\langle E_k \rangle$ self-consistently in the 
model. The lesser and retarded part are all that is needed to formulate the 
self-energy which will be described in detail below.

\subsection{Coulomb matrix elements}
The Coulomb matrix elements are given on the quasi-2D form following \cite{HaugKochBook2004}
\begin{align} \label{Vcoul}
V_{ijkl}(q) = \frac{e^2}{2A \epsilon_0 \epsilon_r } \frac{ F_{ijkl}(q)}{q}
\end{align}
with $\epsilon_r$ being the dielectric constant and $A$ is the system area while the other constants have their usual meaning. The relative constant $\epsilon_r$ is the equilibrium value 
for the main material system with no dopant atoms present. To this we will add the non-equilibrium
contribution as shown in Eq.~\eqref{Wret}, making up the total dielectric constant.
The \textit{form factors} are
expressed as
\begin{align}
F_{ijkl}(q)=\int dz \int dz'
\varphi_{i}^*(z)\varphi_{j}(z) \e^{-q|z-z'|}\varphi_{k}^*(z')\varphi_{l}(z').
\end{align}
where the integral is taken along the growth direction, which defines the quantization direction of the quasi-2D system and $\varphi_i(z)$ are the quantized states in
the $z$-direction used as a basis for the simulation. The numerical integration is
greatly simplified by expressing the wave functions as a
truncated Fourier series~\cite{BonnoJAP2005}. All form factors are then easily calculated
and are thus available at a reasonable computational cost for all $q$. 
These form factors will also be close to exact as the number of 
Fourier components needed
 for the well behaved wave functions of
the heterostructures is limited.   

\subsection{Lesser and retarded expressions for the screened interaction}
We define the SPPA following \cite{HaugKochBook2004} as 
\begin{align} \label{SPPA}
  \frac{1}{\epsilon(\mathbf{q},E)} \approx 1 +
  \frac{E^2_{pl}(q)}{(E+i\delta)^2-E_{eff}^2}
\end{align}
where the effective plasmon pole is chosen to ensure that
the approximation is correct both in the long wavelength $and$ the 
static limit, as thoroughly discussed in \cite{HaugKochBook2004}, 
and given in the two cases of screening dimensions as
\begin{align} \label{Eeff2D}
 \text{2D:} \hspace{1cm}  E_{eff}^2(q)=E_{pl}^2(q)(1+\frac{q}{\kappa})+\nu_q^2 
\end{align}
and 
\begin{align}\label{Eeff3D}
 \text{3D:} \hspace{1cm}  E_{eff}^2(q)=E_{pl}^2(1+\frac{q^2}{\kappa^2})+\nu_q^2 
\end{align}
where the plasma frequency is defined as 
\begin{align*}
 \text{2D:}  \hspace{1cm}   E_{pl}(q) & = \sqrt{\frac{\hbar^2e^2n^{2D}}{2\epsilon_r\epsilon_0 m}q} \\ \nonumber
 \text{3D:} \hspace{1cm}   E_{pl}\phantom{(q)}    &= \sqrt{\frac{\hbar^2e^2n^{3D}}{\epsilon_r\epsilon_0 m}}
\end{align*}
where $n^{2D}$ and $n^{3D}$ are, respectively, the effective 2D and 
3D carrier densities. $m$ is the effective mass of the
electron and $\nu_q^2$ is a term proportional to $q^4$
which should take into account the \textit{pair continuum}~\cite{HaugKochBook2004}. In this 
work we set $\nu_q^2= (\hbar^2 q^2 /2m)^2$. % change 12 feb
Furthermore, $\kappa$ is the inverse of the static screening length given in the two cases as
\begin{align} \nonumber
 \text{2D:} \hspace{1cm}  & \kappa = \frac{me^2}{2\pi\epsilon_0\epsilon_r\hbar^2}f_{k=0} \\ \nonumber
 \text{3D:} \hspace{1cm}  & \kappa = 
\sqrt{\frac{e^2 n^{\rm 3D}}{\epsilon_0\epsilon_r k_BT}}.
\end{align}
In 3D we use the Debye limit of screening and for the 2D case the 
screening becomes a function of the occupation probability $f_k$ at $k=0$ 
for the effective conduction band.

Using the SPPA and the Dirac identity we find for the retarded part of the 
screened interaction
\begin{align} \label{Wret_sppa}
\mathbf{W}^{\rm ret}(\mathbf{q},E) = \mathbf{V}(\mathbf{q}) - \imai \mathbf{V}(\mathbf{q})  \frac{\pi}{2}
\frac{E_{pl}^2}{ E_q} \left[ \delta(E-E_q) -
  \delta(E+E_q ) \right].
\end{align}
This can be linked to the lesser propagator using the 
fluctuation-dissipation theorem as discussed above and it is found to be
\begin{align} \nonumber
\mathbf{W}^<(\mathbf{q},E) &= -\imai f(E) \left[ - 2\Im \{\mathbf{W}^{\rm ret} \} \right] \\  % change 12 feb
 &= -\imai\pi \mathbf{V}(\mathbf{q}) \frac{E_{pl}^2}{ E_q}
\left[ f(E_q) \delta(E-E_q) +
  (f(E_q)+1)\delta(E+E_q) \right].
\end{align}
These are the standard non-interacting bosonic propagators, which are the same
that we have used previously for the optical phonons in our model.
Following the same treatment applied to phonons one can formulate expressions
for the retarded and lesser self-energies. However the static term 
in the retarded interaction has to be taken into account. This is the
exchange shift and it is crucial to preserve the limits of the 
SPPA, static and long wavelength cases.

According to standard Feynmann rules we express the diagram for
the screened interaction self-energy in the GW approximation~\cite{HedinPR1965}
as
\begin{align} \label{SigmaTime}
\Sigma_{\alpha\alpha'}(\mathbf{k},t,t') = \imai \hbar \sum_{\beta\beta'} \sum_\mathbf{k'}
G_{\beta\beta'}(\mathbf{k'},t,t') W_{\alpha\beta\beta'\alpha'}(\mathbf{k-k'},t,t') ,
\end{align}
where the times are now on the Keldysh contour. Using the Langreth rules 
\cite{LangrethBook1976}
we can transform them to the real time axis. Schematically we have for products 
structured as above the following rules: 
\begin{align}
\Sigma^< &= G^< W^< \\
\Sigma^{\rm ret} &= G^{<}W^{\rm ret}+G^{\rm ret}W^{<}+G^{\rm ret}W^{\rm ret}
\end{align}
where the time arguments are implicit and are ordered as in the product 
in Eq.~\eqref{SigmaTime}.
At this point we make a rather drastic approximation and express the 
screened interaction in the SPPA.
The static term in the retarded screened interaction is present 
in the first and third term. In the first it gives rise to the 
exchange described below, however in the third, it does not contribute as
the retarded Green's function, considered as a function of energy, 
only has poles in the lower imaginary half plane. Closing the contour 
in the upper plane prevents contributions from this term.

We can now proceed and formulate the self-energies for the screened 
interaction, and the treatment is analogous to that presented in
Ref.~\cite{WackerPhysRep2002} for optical phonons, 
with the addition of the exchange term treated below. 
Here we make the crucial assumption that the self-energies are 
functions of energy only, in contrast with the more computationally intensive approach of Ref.~\cite{SchmielauAPL2009} that uses full energy and $k$-dependent terms. Here the full $k$-dependence is replaced by an average contribution, by evaluating the scattering matrix elements for a set of typical
values for $E_k$ and $E_ {k'}$, as explained in detail in~\cite{WackerQuantEl2013}. 
This brings the scattering matrix elements out of the
integral over $E_{k'}$ and we write
the lesser and retarded self-energy as 
\begin{multline}
\Sigma^{<}_{\alpha\alpha'}(E)=
\sum_{\beta\beta'}\int \frac{\d \theta}{2\pi} X^{sppa-}_{\alpha\alpha'\beta\beta'}(\theta)
f_{\rm Bose}(E_{eff}(\theta))
\int_0^{\infty}\d E_{k'}G^{<}_{\beta\beta'}(k',E-E_{eff}(\theta))
\\
+\sum_{\beta\beta'}\int \frac{\d \theta}{2\pi}X^{sppa+}_{\alpha\alpha'\beta\beta'}(\theta)
(f_{\rm Bose}(E_{eff}(\theta))+1)
\int_0^{\infty}\d E_{k'}G^{<}_{\beta\beta'}(k',E+E_{eff}(\theta))
\label{EqSigmasppaless}
\end{multline}
and
\begin{equation}\begin{split}
\Sigma^{\rm ret}_{\alpha\alpha'}(E)=&
\sum_{\beta\beta'}\int \frac{\d \theta}{2\pi}X^{sppa-}_{\alpha\alpha'\beta\beta'}(\theta)
(f_{\rm Bose}(E_{eff}(\theta))+1)
\int_0^{\infty}\d E_{k'}G^{{\rm ret}}_{\beta\beta'}(k',E-E_{eff}(\theta))\\
&+
\sum_{\beta\beta'}\int \frac{\d \theta}{2\pi}X^{sppa+}_{\alpha\alpha'\beta\beta'}(\theta)
f_{\rm Bose}(E_{eff}(\theta))  
\int_0^{\infty}\d E_{k'}G^{{\rm ret}}_{\beta\beta'}(k',E+E_{eff}(\theta))\\
&+\frac{1}{2}
\sum_{\beta\beta'}\int \frac{\d \theta}{2\pi}X^{sppa-}_{\alpha\alpha'\beta\beta'}(\theta)
\int_0^{\infty}\d E_{k'}G^{<}_{\beta\beta'}(k',E-E_{eff}(\theta))\\
&-\frac{1}{2}
\sum_{\beta\beta'}\int \frac{\d \theta}{2\pi}X^{sppa+}_{\alpha\alpha'\beta\beta'}(\theta)
\int_0^{\infty}\d E_{k'}G^{<}_{\beta\beta'}(k',E+E_{eff}(\theta))\\
&+\imai\int\frac{\d E''}{2\pi}
{\cal P}\left\{\frac{1}{E''}\right\}
\Big[\sum_{\beta\beta'}\int \frac{\d \theta}{2\pi}X^{sppa-}_{\alpha\alpha'\beta\beta'}(\theta)
\int_0^{\infty}\d E_{k'}G^{<}_{\beta\beta'}({\bf k}',E-E_{eff}(\theta)-E'')\\
&\phantom{+\imai\int\frac{\d E''}{2\pi}{\cal P}\left\{
\frac{1}{E''}\right\}}
-\sum_{\beta\beta'}X^{sppa+}_{\alpha\alpha'\beta\beta'}(\theta)
\int_0^{\infty}\d E_{k'}G^{<}_{\beta\beta'}({\bf k}',E+E_{eff}(\theta)-E'')\Big]
\label{EqSigmaspparet}
\end{split}\end{equation}
where the last term contains an principal value integral. As this term involves yet
another integral in energy space, it is of higher numerical complexity than the other
terms and currently neglected to expedite the numerical calculations.
The factor
$f_{\rm Bose}(E_{eff})=1/(\exp(E_{eff}/k_BT)-1)$ is again the Bose
distribution where the electron temperature enters. 
In each term there is an average over all possible scattering angles.
The scattering matrix elements are given as
\begin{align} \label{xsppa}
  X^{sppa}_{\alpha\alpha'\beta\beta'}(\theta)=\frac{1}{2\pi}
  \frac{me^2}{2\epsilon_0\epsilon_r \hbar^2} \frac{F_{\alpha\alpha'\beta\beta'}(\theta)}{q(\theta)} \frac{E_{pl}(q(\theta))}{2E_{eff}(q(\theta))}
\end{align}
and the plus and minus variants used above comes from the choice of 
typical values for the evaluation. In our $k$-independent approximation we put
\begin{equation}
X^{{\rm sppa}\pm}_{\alpha\alpha',\beta\beta'}
=X^{\rm sppa}_{\alpha\alpha',\beta\beta'}(E_{typ},E_{typ}
+|\Delta E \mp E_{\rm eff}|)
\end{equation}
with $\Delta E$ takes the level difference properly into account for
all combinations of indices. $E_{typ}$ is a representative value chosen
to give a similar scattering rate compared to the case of thermalised 
subbands at the given lattice temperature.

\subsection{Exchange shift}
As the Hartree-Fock self-energies are local in time they depend only on the
difference in coordinates, and the exchange term is normally evaluated 
directly from the diagrammatic rules to be
\begin{align}
  \Sigma^x_{\alpha\alpha'}(\mathbf{k},t_1,t_2) = \hbar  \sum_{\beta\beta'} \sum_{\mathbf{q}}  
V_{\alpha\beta\beta'\alpha'}(|\mathbf{ q}|) \delta(t_1,t_2) \imai G_{\beta\beta'}(\mathbf{k-q},t_1,t_2)
\end{align} 
which simplifies if only the difference in time matters and we 
find
\begin{align} \nonumber
\Sigma^x_{\alpha\alpha'}(\mathbf{k},E) &= - \sum_{\beta\beta'} \sum_{\mathbf{ q}} 
\int \d \tau \e^{\imai E/\hbar \tau } 
V_{\alpha\beta\beta'\alpha'} (|\mathbf{q}|) \delta(\tau)
\int \frac{\d E'}{2\pi \imai} \e^{-\imai E'/\hbar \tau} 
G^<_{\beta\beta'}(\mathbf{k-q},E') \\ \nonumber
 &=  -\sum_{\beta\beta'} \sum_{\mathbf{q} } V_{\alpha\beta\beta'\alpha'}(|\mathbf{q}|) 
\int \frac{\d E'}{2\pi\imai} G^<_{\beta\beta'}(\mathbf{k-q},E') \\ \nonumber 
%& = - \sum_{\beta\beta'} \sum_{\mathbf{ q}}  V_{\alpha\beta\beta'\alpha'}(|\mathbf{ q}|) 
%\rho_{\beta\beta'}(\mathbf{ k-q}) \\ \nonumber
 & = - \sum_{\beta\beta'} \sum_{\mathbf{ k'}}  V_{\alpha\beta\beta'\alpha'}(|\mathbf{ k-k'}|)
   \rho_{\beta\beta'}(\mathbf{ k'}) \\ \nonumber
 & \approx - \sum_{\beta\beta'} \bar{V}_{\alpha\beta\beta'\alpha'}(E_{k},E_{k'}) 
\sum_{\mathbf{ k'} } \rho_{\beta\beta'}(\mathbf{ k'}) \\ 
 & = - \sum_{\beta\beta'} \bar{V}_{\alpha\beta\beta'\alpha'}(E_{k},E_{k'}) 
\frac{A}{2} \bar{\rho_{\beta\beta'}}
\end{align} 
where we moved the Coulomb matrix elements out of the sum over
$\mathbf{k}$ in order to do  the sum and reach the final 2D sheet
densities $\bar{\rho}_{\beta\beta'}$. The factor of 1/2 appears as we do not
sum over spin for the exchange self-energy. The $\bar{V}(E_{k},E_{k'})$ is
now an angle averaged quantity over the possible angles between
$\mathbf{k}$ and $\mathbf{k'}$, where we again use typical values 
for $E_k$ and $E_{k'}$.

\section{Results}
\begin{figure}
\centering
\includegraphics[width=0.40\textwidth]{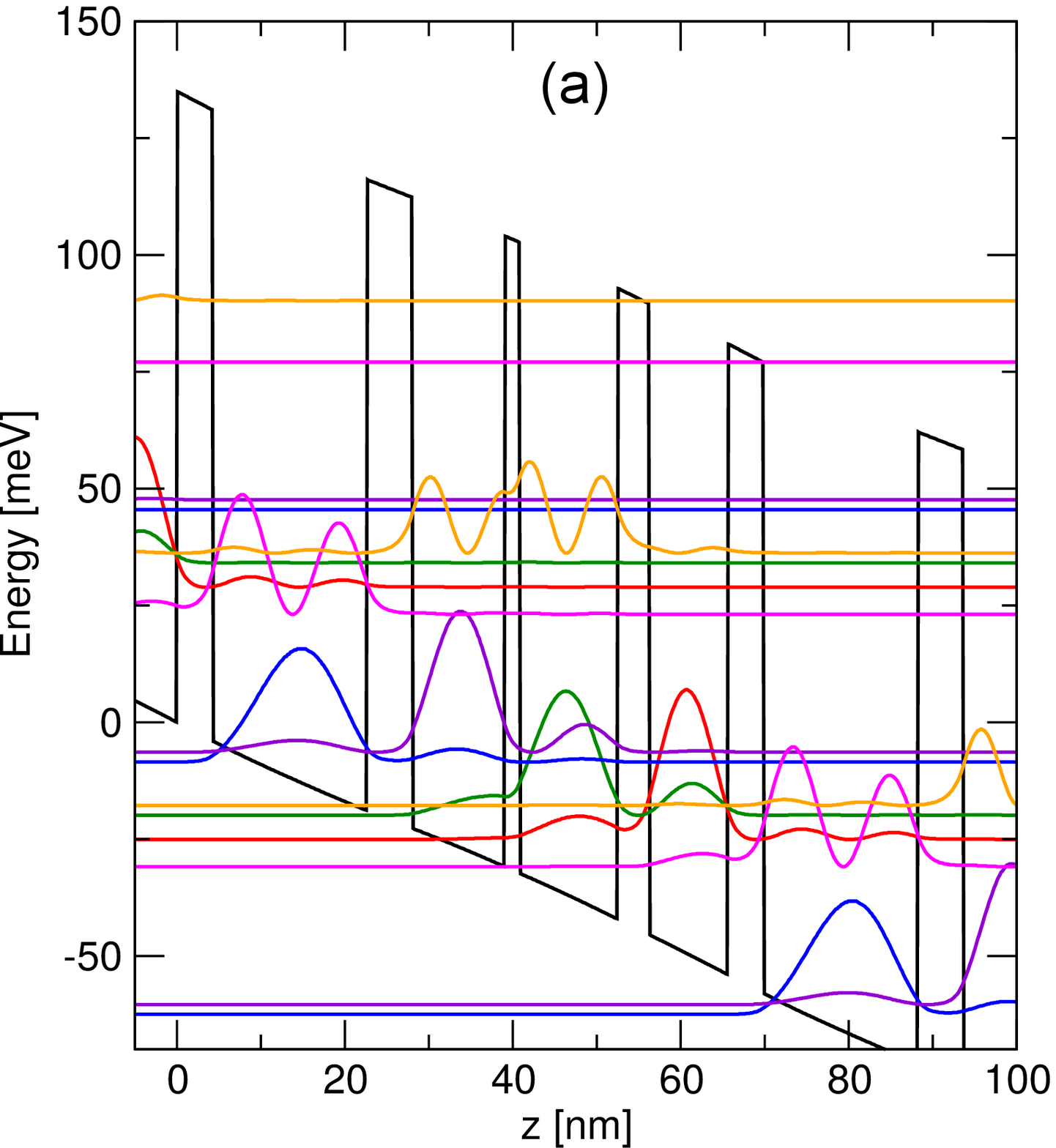}
\includegraphics[width=0.348\textwidth]{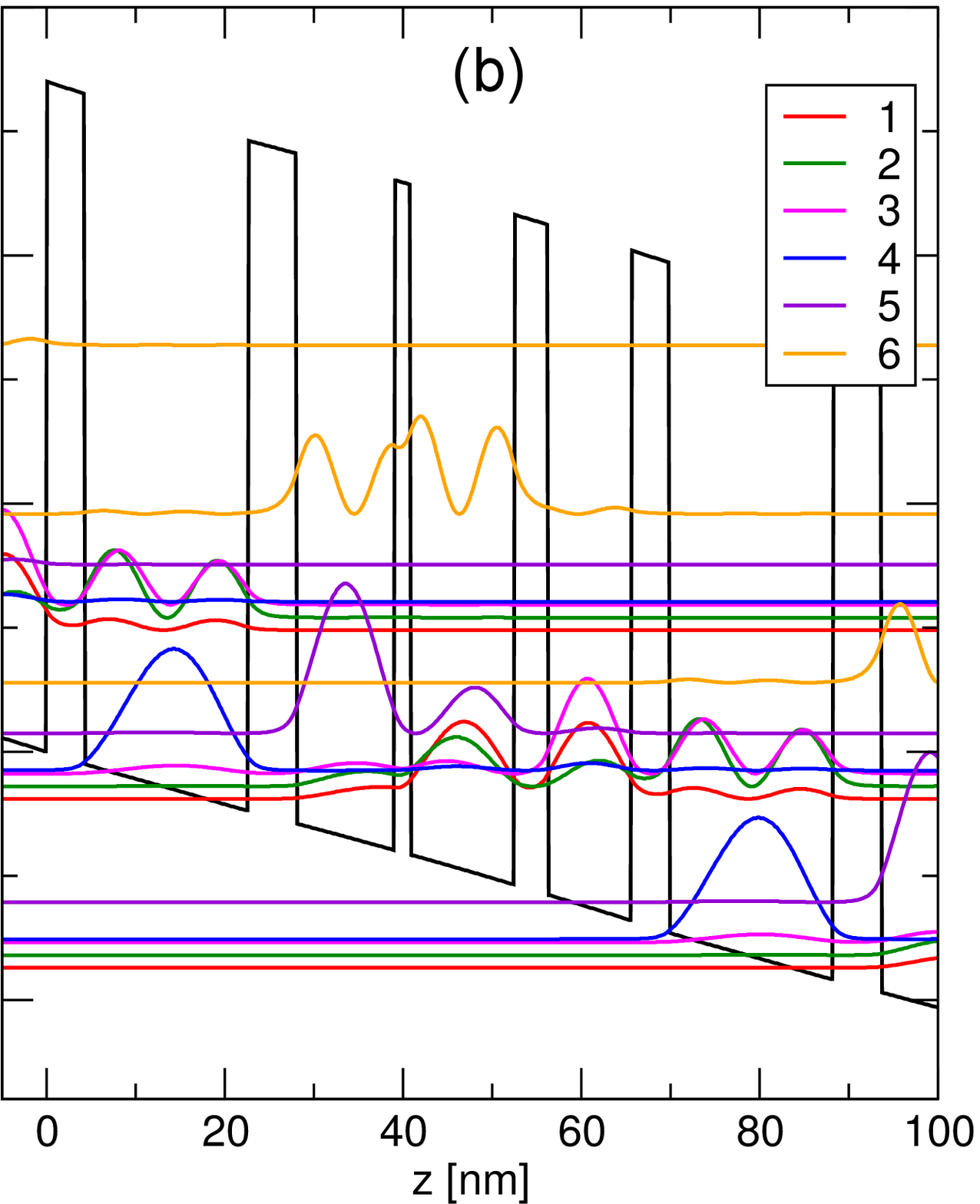}
\caption{Band diagram of a four-well QCL from Ref.~\cite{AmantiNJP2009}. Here the conduction band
edge is shown as solid black lines spanning a periodic sequence of quantum wells typical for the
QCL. The barrier material is ${\rm Al}_{0.15}{\rm Ga}_{0.85}{\rm As}$ while the wells are pure GaAs. 
In the widest well dopant atoms are placed in order to n-dope the structure, giving rise to 
impurity scattering, whereas interface roughness scattering originates from imperfect material 
interfaces. Probability densities are shown for the most relevant states. This design
 employs resonant tunneling injection from 4 to 5, building up inversion between 5 and 2.
The lower laser state 2 is efficiently depopulated via resonant tunneling out to 3 via 1, 
combined with phonon extraction to the starting point; state 4, shifted by one period.
In (a) the alignment at design bias is shown, at a bias of 54 mV per period. 
In (b) the alignment at the parasitic resonance at 34 mV is 
shown, giving rise to the pre-peak in Fig.~\ref{FigCurr}.}
\label{FigBands}
\end{figure}

Our model, without electron-electron scattering has been previously tested successfully on superlattices
and QCL heterostructures~\cite{WackerQuantEl2013,FranckieOE2015,WingeOE2014}. 
The case considered here is a QCL design from 2009 \cite{AmantiNJP2009}. 
The periodic structure is displayed in Fig.~\ref{FigBands} together with the 
probability density for the subbands relevant for transport. These type
of four-well THz QCL structures have previously been troublesome to simulate, see for 
example~\cite{WingeAPL2012}, due to a pre-peak
feature which is not recovered experimentally, at least not to the same extent
as in the simulations. In theory, a lot more weight is put on the pre-peak at
the cost of a lower main peak, compared to the opposite situation in experiments.

This design was also tested experimentally for the different 
doping concentrations, which makes it ideal as we want to study the impact 
of electron-electron interaction. The experimental findings are summarized 
in Table~\ref{ExpTable}. Sample A is doped to acquire a sheet density normal
for THz QCLs while sample B has about a factor three higher doping concentration.
For the SPPA, there is clearly an issue with 
vanishing carrier sheet density. In such a situation the physical picture 
of a continuous spectra of modes of collective oscillations is questionable as
a valid representation of the actual physics. 

\begin{figure}
\centering
\includegraphics[width=0.6\textwidth]{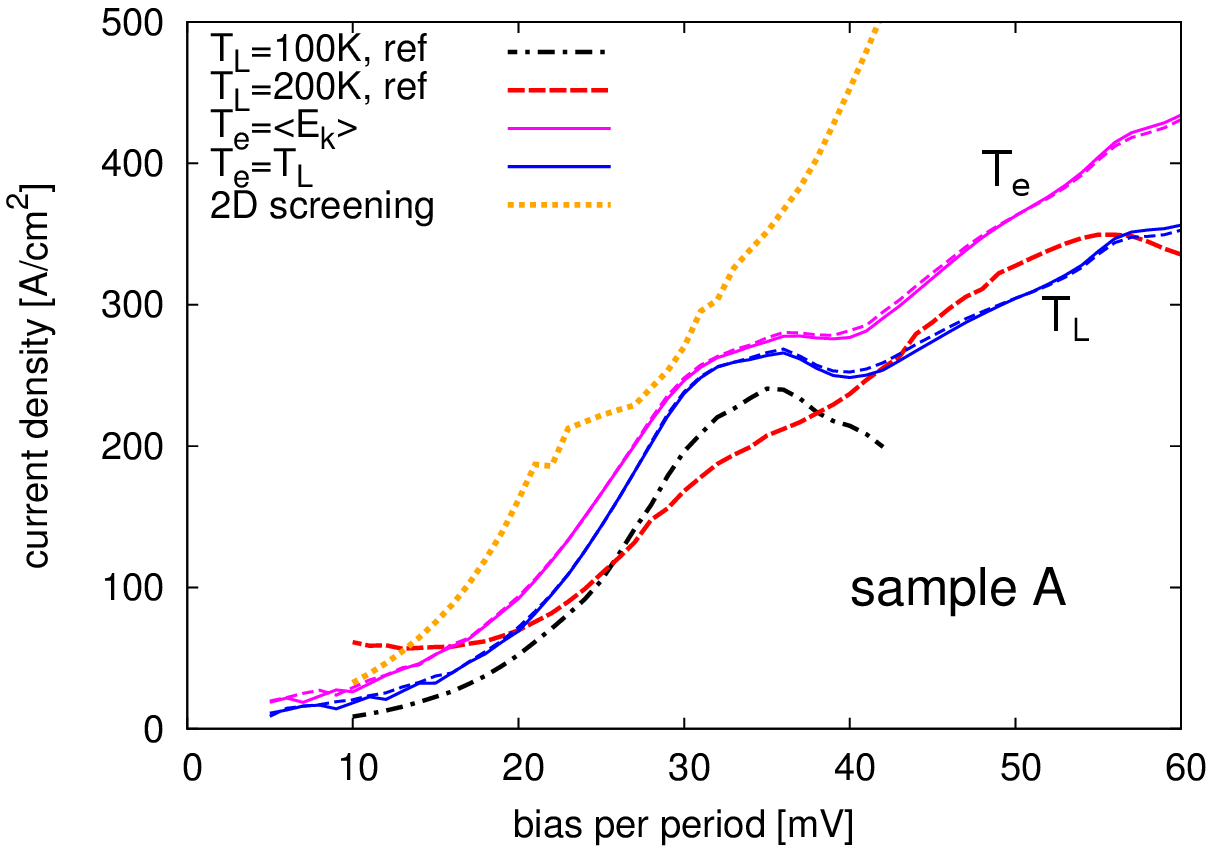}
\includegraphics[width=0.6\textwidth]{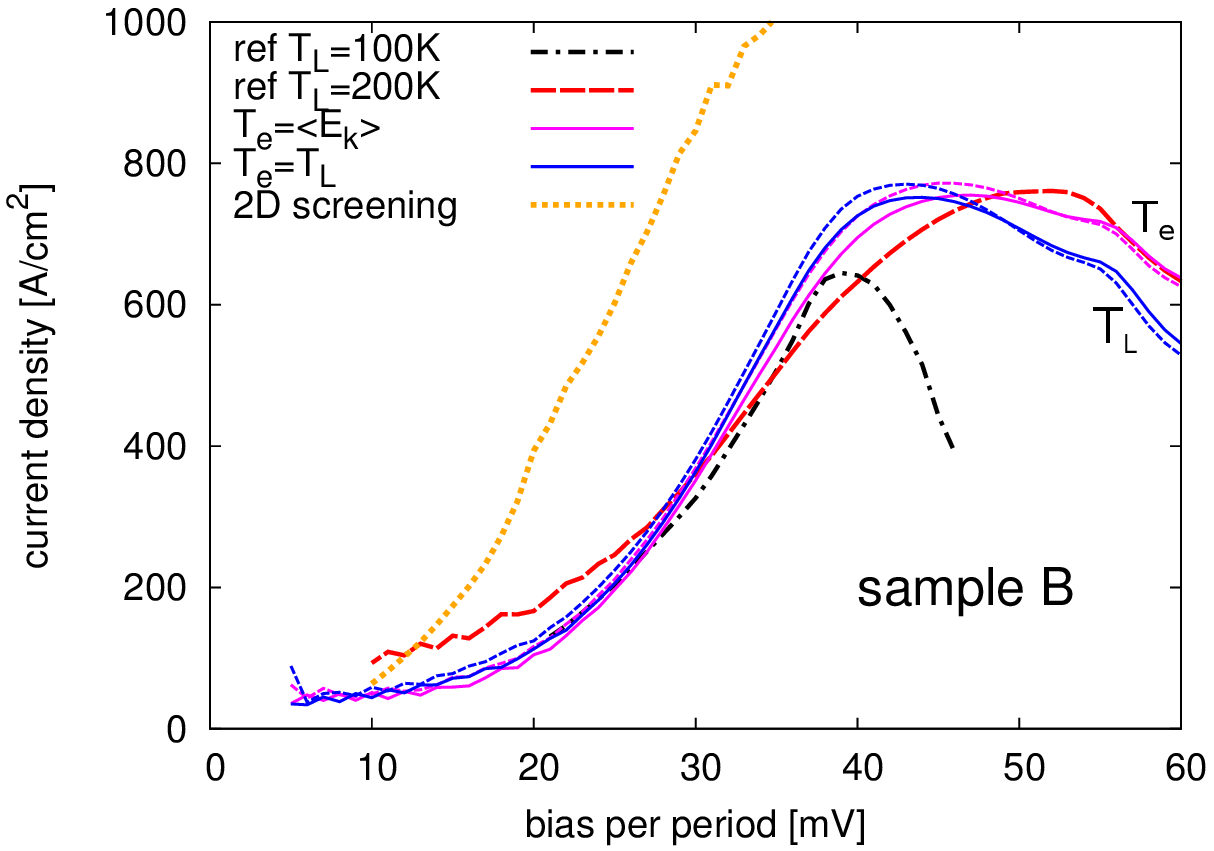}
\caption{Current simulations for sample A and B in the respective plots. 
Reference simulations are shown as black (dot-dashed) and red (long-dashed) lines for 100 K and 200 K 
lattice temperature, respectively. Simulations with the SPPA self-energy at 100 K, are
shown as blue and magenta lines for 3D screening using Eq.~\eqref{Eeff3D} and orange (square-dotted) lines 
for 2D screening using Eq.~\eqref{Eeff2D}. Simulations without the exchange shift is indicated
by short-dashed lines.}
\label{FigCurr}
\end{figure}

The results of adding the new self-energy to the model is displayed for both
sample A and B in Fig.~\ref{FigCurr}. As a reference we show currents calculated 
without the SPPA self-energy for both 100 and 200 K lattice temperature.
For high temperature the current increases almost linearly up to design bias, 
around 54 mV as depicted in Fig.~\ref{FigBands}(a), however for low temperature 
the current enters a region of Negative Differential
Resistance (NDR) after $\sim 38$ mV for both samples. After entering this region it is no longer
possible to converge the simulations; that is why the black line abruptly stops. 
The lack of convergence is attributed to artificially low lifetimes of the states due
to the lack of important scattering mechanisms. In order to reach convergence
in the self-consistent process, the energy resolution
would have to be increased significantly. This is currently being investigated but the 
higher numerical complexity prevents us from showing any results here.

The pre-peak feature occurs at a low bias, before there is sufficient 
gain to start laser operation and it indicates that some important scattering
is missing in the model as this is not the behavior seen in experiments.
For sample A, a threshold current of 175 A/cm$^2$ is observed for 10 K heatsink temperature
and this gradually increases to 430 A/cm$^2$ for the maximum operating temperature.
The structure should thus be able to reach design bias also for low temperature.
The NDR in both samples occurs close to the bias
matching a potential drop of the energy of one
optical phonon, which is about 36 meV for AlGaAs/GaAs systems. 
This strongly favors the parasitic current channel shown in Fig.~\ref{FigBands}(b).

When the SPPA self-energy is included in the low temperature simulations the 
current changes drastically, and
employing a 2D or 3D screening model makes a big difference. The orange lines in
Fig.~\ref{FigCurr} shows simulations with 2D screening at a lattice temperature of 100K, 
and the current
continues off the scale and is much too large compared to the experimental
data. The reason for this is that the plasmon
dispersion has a finite value for the $q\to0$ limit for 3D but tends to 
zero in the 2D case, as shown in Eqs.~(\ref{Eeff2D}-\ref{Eeff3D}). 
This effectively gives additional screening for the 
3D case as the plasmon quanta, limited downwards by the plasma energy, 
always have to be exchanged with the collective modes that are assumed 
to be in thermal equilibrium. For the 2D case, there is nothing limiting
the system from exchanging plasmons with vanishing energy, making both
the occupation factor in Eqs.~(\ref{EqSigmasppaless}-\ref{EqSigmaspparet})
and the Coulomb matrix element tend to large values. 
In the long wavelength ($q\to0$) limit the system should tend to a 3D electron gas,
which is respected by the 3D but not the 2D screening model. This, together
with the observation that the 2D screening model gives an unrealistic current 
contribution, is the reason for focusing the rest of our discussion on the 3D screening model.

A few different cases are displayed in Fig.~\ref{FigCurr} for the simulations with the SPPA 
self-energy with 3D screening at $T_L=100$ K. Qualitatively we observe better agreement as the  
simulations are similar to the reference case at 200 K, which is expected. The quantative
agreement is shown in Table~\ref{ExpTable}.
Here we use a lattice temperature 30 K higher than the \textit{heatsink}
temperature reported in the experiment. No laser field is included, only off-state simulations
are shown. The agreement is reasonable except for the $J_{\rm NDR}(10 {\rm K})$ for sample 
A. For sample B the agreement appears to be better but as we consider off-state current and 
on-state current probably is higher, this might be too large as well.
The experimental decrease in current with lattice temperature is thus not seen in the simulations.
The way we interpret this is that the possibility of exchanging plasmons with the environment
opens up new transport channels for the electrons, and that this effect
actually is larger than the hypothesized effect of increased dephasing
which would decrease the tunneling currents, such as the parasitic 
resonance at the NDR peak. This effect plays a larger role at low temperature
where phonon scattering is weak.

In Fig.~\ref{FigCurr} different limits of
plasmon-phonon coupling are shown, indicated by $T_L$ and $T_e$.
In case of a strong coupling of the plasmon and phonon
baths the plasmons would be quickly cooled by the phonons. This 
is shown as blue lines where the electron temperature has been set equal to
the lattice temperature $T_L$. The opposite would be a weak coupling, and here
the plasmons are only coupled indirectly through the mean electron temperature
calculated as $T_e=\langle E_k \rangle/k_B$ with respect to the bottom of each subband.
This is shown as magenta lines in Fig.~\ref{FigCurr}. 
Due to the occupation factors in Eqs.~(\ref{EqSigmasppaless}-\ref{EqSigmaspparet})
a higher electron temperature increases the strength of the SPPA self-energy.
The impact of switching the exchange self-energy off is shown by the
dashed line of the respective color. 

\begin{figure}
\centering
\includegraphics[width=0.5\textwidth]{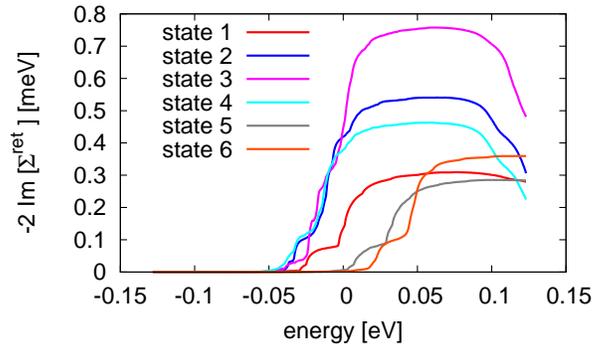}
\caption{Here the imaginary parts of the diagonal elements of the retarded SPPA self-energy is shown for 
simulations at design bias. The state indexing follows the one in Fig.~\ref{FigBands}.
Compared to other scattering mechanisms, the SPPA gives a small contribution to the 
linewidth: about 7\% of the total at this bias. The lattice temperature is 100 K.}
\label{FigSigRet}
\end{figure}

In Fig.~\ref{FigSigRet} the imaginary part
of the retarded self-energy is plotted as a function of energy. This
is directly related to the linewidths of the single particle states. One
finds that the effect of this self-energy on the linewidth is below one meV, 
for all the states.

\begin{table}
  \centering
  \begin{tabular}{l l c c c c}
    &Sample & $\rho^{\rm 2D}$ & $J_{\rm NDR}$(10 K) & $J_{\rm max}$($\sim 150$ K) \\
    \hline
  \multirow{2}{*}{Exp.} &  A(EV1157) & $3.7\times10^{10}$ cm$^{-2}$ & 225 A cm$^{-2}$ & 430 A cm$^{-2}$ \\
  & B(N907) & $1.1\times10^{11}$ cm$^{-2}$ & 810 A cm$^{-2}$ & 920 A cm$^{-2}$ \\
  \hline
   \multirow{2}{*}{Theory} &  A & $3.7\times10^{10}$ cm$^{-2}$ & 465 A cm$^{-2}$ & 460 A cm$^{-2}$ \\
  & B & $1.1\times10^{11}$ cm$^{-2}$ & 710 A cm$^{-2}$ & 800 A cm$^{-2}$ \\
    \hline \hline
  \end{tabular}
  \caption{Experimental results from Ref.~\cite{AmantiNJP2009} compared to simulations.
    Here we list, from left to right, sample name, sheet density, current at the
    NDR feature where lasing stops at low temperature and lastly the maximum
    current at the temperature, observed in experiments, where the laser ceases to operate. }
  \label{ExpTable}
\end{table}

\section{Conclusions}
The screened electron-electron interaction has been included in a crude 
GW approximation via the SPPA, including dynamical screening. 
This leads to increased scattering in our model and a behavior that
better represents the data from experimental studies, showing that these
scattering mechanisms are indeed an important source of inelastic scattering
when the other mechanisms of this type, such as acoustic phonon scattering, 
are weak. 

Exchange effects are shown to play a minor role only at higher doping concentrations, while the
choice of coupling of the plasmon and phonon baths has a larger impact.
We observe that a 3D screening model gives reasonable results due to the finite
value of the plasmon energy in the low $q$ limit, and we conclude that it also
preserves the long wavelength limit.
In contrast, the 2D screening model largely overestimates current and does not 
respect the long wavelength limit.

\ack
Financial support from the COST Action MP1204 in order to execute Short Term Scientific 
missions is thankfully acknowledged, as is support from the Swedish Research Council (VR).

%\bibliographystyle{./bibtex-iopart-num/iopart-num}
%\bibliography{refs_QuantTrans}

\section*{References}
\begin{thebibliography}{10}
\expandafter\ifx\csname url\endcsname\relax
  \def\url#1{{\tt #1}}\fi
\expandafter\ifx\csname urlprefix\endcsname\relax\def\urlprefix{URL }\fi
\providecommand{\eprint}[2][]{\url{#2}}

\bibitem{FaistScience1994}
Faist J, Capasso F, Sivco D~L, Sirtori C, Hutchinson A~L and Cho A~Y 1994 {\em
  Science\/} {\bf 264} 553--556

\bibitem{WilliamsNatPhotonics2007}
Williams B~S 2007 {\em Nat.~Photonics\/} {\bf 1} 517

\bibitem{BelkinPhysScripta2015}
Belkin M~A and Capasso F 2015 {\em Phys. Scripta\/} {\bf 90} 118002

\bibitem{KhanalJOpt2014}
Khanal S, Zhao L, Reno J~L and Kumar S 2014 {\em J.~Opt.\/} {\bf 16} 094001

\bibitem{JirauschekApplPhysRev2014}
Jirauschek C and Kubis T 2014 {\em Appl. Phys. Rev.\/} {\bf 1} 011307

\bibitem{CallebautAPL2004}
Callebaut H, Kumar S, Williams B~S, Hu Q and Reno J~L 2004 {\em
  Appl.~Phys.~Lett.\/} {\bf 84} 645

\bibitem{ManentiJComputElectron2003}
Manenti M, Compagnone F, Di~Carlo A and Lugli P 2003 {\em
  J.~Comput.~Electron.\/} {\bf 2} 433--437

\bibitem{JirauschekPSSC2008}
Jirauschek C and Lugli P 2008 {\em phys.~stat.~sol. (c)\/} {\bf 5} 221--224

\bibitem{LundqvistPhysKondensMater1967}
Lundqvist B 1967 {\em Phys. Kondens. Mater.\/} {\bf 6} 193--205

\bibitem{AndoRMP1982}
Ando T, Fowler A~B and Stern F 1982 {\em Rev.~Mod.~Phys.\/} {\bf 54} 437--672

\bibitem{SchmielauAPL2009}
Schmielau T and Pereira M 2009 {\em Appl.~Phys.~Lett.\/} {\bf 95} 231111

\bibitem{WackerQuantEl2013}
Wacker A, Lindskog M and Winge D 2013 {\em Sel. Top. in Quantum Electron., IEEE
  Journal of\/} {\bf 19} 1200611

\bibitem{KubisPRB2009}
Kubis T, Yeh C, Vogl P, Benz A, Fasching G and Deutsch C 2009 {\em
  Phys.~Rev.~B\/} {\bf 79} 195323

\bibitem{KeldyshSovPhysJETP1965}
Keldysh L~V 1965 {\em Sov.~Phys.~JETP\/} {\bf 20} 1018 [Zh.~Eksp.~Theor.~Fiz.\
  {\bf 47}, 1515 (1964)]

\bibitem{KadanoffBaymBook1962}
Kadanoff L~P and Baym G 1962 {\em Quantum Statistical Mechanics\/} (New York:
  Benjamin)

\bibitem{WackerPhysRep2002}
Wacker A 2002 {\em Phys.~Rep.\/} {\bf 357} 1

\bibitem{HaugKochBook2004}
Haug H and Koch S 2004 {\em Quantum theory of the optical and electronic
  properties of semiconductors\/} (Singapore: World Scientific)

\bibitem{BonnoJAP2005}
Bonno O, Thobel J and Dessenne F 2005 {\em J.~Appl.~Phys.\/} {\bf 97} 043702

\bibitem{HedinPR1965}
Hedin L 1965 {\em Phys.~Rev.\/} {\bf 139} A796

\bibitem{LangrethBook1976}
Langreth D~C 1976 {\em Linear and Nonlinear Electron Transport in Solids\/} ed
  Devreese J~T and van Doren V~E (New York: Plenum Press)

\bibitem{FranckieOE2015}
Francki\'{e} M, Winge D~O, Wolf J, Liverini V, Dupont E, Trinit\'{e} V, Faist J
  and Wacker A 2015 {\em Opt. Express\/} {\bf 23} 5201--5212

\bibitem{WingeOE2014}
Winge D~O, Lindskog M and Wacker A 2014 {\em Opt. Express\/} {\bf 22}
  18389--18400

\bibitem{AmantiNJP2009}
Amanti M~I, Scalari G, Terazzi R, Fischer M, Beck M, Faist J, Rudra A, Gallo P
  and Kapon E 2009 {\em New~J.~Phys.\/} {\bf 11} 125022

\bibitem{WingeAPL2012}
Winge D~O, Lindskog M and Wacker A 2012 {\em Appl.~Phys.~Lett.\/} {\bf 101}
  211113

\end{thebibliography}

\end{document}